\documentclass[aps,prd,10pt,notitlepage,nofootinbib]{revtex4-1}

\usepackage[utf8]{inputenc}
\usepackage{amsmath,amssymb,amsfonts}
\usepackage{newtxtext,newtxmath}
\usepackage{mathtools}

\usepackage{bbold}
\usepackage{bm}
\usepackage{graphicx}
\usepackage[usenames,dvipsnames]{xcolor}
\usepackage{color}
\usepackage[colorlinks=true,linkcolor=blue,urlcolor=blue,citecolor=blue]{hyperref}

\usepackage{slashed}
\usepackage[english]{babel}
\usepackage{dcolumn}
\usepackage{pifont}
\usepackage{dsfont,mathrsfs}
\usepackage{cancel}
\usepackage{bigints}
\usepackage{accents}
\usepackage{soul}
\usepackage{multirow}

\usepackage{amsmath, amssymb}
\usepackage{bbold}
\usepackage{makecell}
\usepackage{simpler-wick}

\usepackage{orcidlink} 

\newcommand{\nn}{\nonumber}

\newcommand{\FB}[1]{\left(#1\right)}

\newcommand{\TB}[1]{\left[#1\right]}
\newcommand{\AB}[1]{\left<#1\right>}

\newcommand{\munu}{{\mu\nu}}

\newcommand{\del}{\partial}

\newcommand{\qtil}{\tilde{q}_\perp}
\newcommand{\kp}{k_\perp}
\newcommand{\kpv}{\bm{k}_\perp}
\newcommand{\qp}{q_\perp}
\newcommand{\pp}{p_\perp}
\newcommand{\qpv}{\bm{q}_\perp}
\newcommand{\ppv}{\bm{p}_\perp}

\newcommand{\floor}[1]{\lfloor #1 \rfloor}

\allowdisplaybreaks

\begin{document}
\title{Evaluation of a Double Integral with Exponentials and Two Generalized Laguerre Polynomials and Its Associated Moments}
\author{Snigdha Ghosh\orcidlink{0000-0002-2496-2007}}
\affiliation{Government General Degree College Kharagpur-II, Madpur, Paschim Medinipur - 721149, West Bengal, India}	
\email{snigdha.ghosh@bangla.gov.in}
\email{snigdha@ggdckharagpur2.ac.in}
\email{snigdha.physics@gmail.com}
%
%
	
\begin{abstract}
	\section*{Abstract} \vspace*{-0.5cm}
We analytically evaluate in terms of finite sums, the following definite improper integral, which appears in the quantum field theoretical calculation involving elementary charged particles having charge $\pm e$ in presence of a constant external magnetic field $\bm{B}=B\hat{\bm{z}}$
$$
I_{abln}^{\mu_1\dots\mu_r}(\bm{q}_\perp,|eB|) = \int\!\!\dfrac{d^2k_\perp}{(2\pi)^2}k_\perp^{\mu_1}  \dots k_\perp^{\mu_r} \exp \bigg(-\frac{\bm{k}_\perp^2+\bm{p}_\perp^2}{|eB|}\bigg) L^a_l \bigg(\frac{2\bm{k}_\perp^2}{|eB|}\bigg) L^b_n \bigg(\frac{2\bm{p}_\perp^2}{|eB|}\bigg)\bigg|_{\bm{p}_\perp=(\bm{q}_\perp+\bm{k}_\perp)}
$$
where $\mu_i \in \{1,2\}$, $l\ge 0$, $n\ge 0$ and $L_l^a(x)$ are the generalized Laguerre polynomials. We also provide explicit expressions of the result up to the fourth moment i.e. $r=4$, which are sufficient for most of the field theoretical calculations. 
\end{abstract}

\maketitle

\section{Introduction}
The study of dynamics of elementary charged particles (having electric charge $\pm e$) in presence of an constant external magnetic field $\bm{B} = B \hat{\bm{z}}$ in the framework of quantum field theory (QFT) often requires evaluation of non-trivial improper integrals and sums that involve generalized Lagueree polynomials (see Refs.~\cite{Andersen:2014xxa,Miransky:2015ava} for reviews and the references therein). In particular, while calculating Feynman diagrams containing two charged particles in a loop at finite $B$, one usually encounters the following type of integral~\cite{Wang:2020dsr,Das:2021fma,Ghosh:2018xhh,Ghosh:2019fet,Ghosh:2020wqx}
 \begin{eqnarray}
 	I_{abln}^{\mu_1\dots\mu_r} = \int\!\!\dfrac{d^2k_\perp}{(2\pi)^2}k_\perp^{\mu_1}  \dots k_\perp^{\mu_r}
 	\exp \bigg(-\frac{\kpv^2+\ppv^2}{|eB|}\bigg) L^a_l \bigg(\frac{2\kpv^2}{|eB|}\bigg) L^b_n \bigg(\frac{2\ppv^2}{|eB|}\bigg)\bigg|_{\ppv=(\qpv+\kpv)} \label{I1}
 \end{eqnarray}
 where $\mu_i \in \{1,2\}$, $l\ge 0$, $n\ge 0$, $L_l^a(x)$ are the generalized Laguerre polynomials. For $\qpv = \bm{0}$, the above integral can be trivially evaluated using the orthogonality of the generalized Laguerre polynomials and their recurrence relations~\cite{Ghosh:2020wqx,Chakraborty:2017vvg,GRADSHTEYN1980904}. However, for $\qpv \ne 0$, the integral is non-trivial to evaluate analytically and one often resort to numerical calculations~\cite{Mondal:2023vzx}. The numerical evaluation of integral of the form~\eqref{I1} becomes uncertain for extreme values of the ratio $\frac{\qpv^2}{|eB|}$ or when $|l-n|$ becomes large~\cite{Chakraborty:2017vvg}. In Refs~\cite{Chakraborty:2017vvg,Das:2021fma,Wang:2020dsr}, some special cases of the integral defined in Eq.~\eqref{I1} are analytically evaluated using different methodologies. In this work, we follow the method discussed in Ref.~\cite{Das:2021fma} and generalize it to analytically evaluate the integral in Eq.~\eqref{I1}
 
 The article is organized as follows. In the next section we mention the required notation and convention as used in filed theoretical calculations. Next, in Sec.~\ref{sec.evaluation}, we discuss the analytical evaluation of the integral. Finally, in Sec.~\ref{sec.moments}, we provide the explicit expressions of the results up to the fourth moment (i.e. $r=4$).
 
\section{Notations and Convention} \label{sec.notation}
In the quantum field theoretical calculations, we mostly use four dimensional flat Minkowski space-time having metric tensors $g^\munu = g_\munu = \text{diag}(1,-1,-1,-1)$. The contravariant four vector $k^\mu$ have components $ k^\mu \equiv (k^0,\bm{k})$. The corresponding covariant vector $k_\mu = g_\munu k^\nu \equiv (k^0,-\bm{k})$. In the presence of external magnetic field $\bm{B} = B\hat{\bm{z}}$, the rotational symmetry of the system is broken; and with respect to the direction of magnetic field, we decompose $k^\mu=(k_\parallel^\mu + k_\perp^\mu)$ and $g^\munu = (g^\munu_\parallel+g^\munu_\perp)$ where $g^\munu_\parallel = \text{diag}(1,0,0,-1)$ and $g^\munu_\perp = \text{diag}(0,-1,-1,0)$. Thus, the parallel and perpendicular vectors have components $k_\parallel^\mu = g_\parallel^\munu k_\nu = (k^0,0,0,k_z)$  and $k_\perp^\mu = g_\perp^\munu k_\nu = (0,k_x,k_y,0)$. Accordingly the norm of the four vector $k^2 \equiv (k_0^2-\bm{k}^2) = (k_0^2-k_x^2-k_y^2-k_z^2)$ decomposes as $k^2= (k_\parallel^2+k_\perp^2)$, where $k_\parallel^2 = (k_0^2-k_z^2)$ and $k_\perp^2=-(k_x^2+k_y^2)$. We note that unless $k_\perp$ is null, it is a space-like vector and has a negative norm $k_\perp^2<0$. We also define $\bm{k}_\perp^2 = -k_\perp^2 = (k_x^2+k_y^2)\ge0$.

\section{Analytic Evaluation of the Integral} \label{sec.evaluation}
In order to evaluate the integral defined in Eq.~\eqref{I1}, we first multiply both sides of Eq.~\eqref{I1} by $t^lu^n$ where $(t,u) \in \mathbb{R}$, $|t|\ll1$ and $|u|\ll 1$. This is then followed by taking a sum over $l$ and $n$ which yields
\begin{eqnarray}
	\sum_{l=0}^{\infty} \sum_{n=0}^{\infty} I_{abln}^{\mu_1\dots\mu_r}t^lu^n = \int\!\!\dfrac{d^2k_\perp}{(2\pi)^2}k_\perp^{\mu_1} \dots k_\perp^{\mu_r}
	\exp \bigg(-\frac{\kpv^2+\ppv^2}{|eB|}\bigg)   \sum_{l=0}^{\infty} L^a_l \bigg(\frac{2\kpv^2}{|eB|}\bigg)t^l \sum_{n=0}^{\infty} L^b_n \bigg(\frac{2\ppv^2}{|eB|}\bigg)u^n \bigg|_{\ppv=(\qpv+\kpv)}. \label{I2}
\end{eqnarray}
We now use the generating function of the Laguerre Polynomials~\cite{GRADSHTEYN1980904} 
\begin{eqnarray}
\sum_{l=0}^{\infty} L_l^a(x)z^n = 	\frac{1}{(1-z)^{a+1}} \exp\FB{\frac{zx}{z-1}} ~~~~\text{for}~ |z|<1	
\end{eqnarray}
 and evaluate the sums on the rhs of Eq.~\eqref{I2} to obtain
\begin{eqnarray}
\sum_{l,n=0}^{\infty}  I_{abln}^{\mu_1\dots\mu_r}t^lu^n &=&	\frac{1}{(1-t)^{a+1}(1-u)^{b+1}}  \int\!\!\dfrac{d^2k_\perp}{(2\pi)^2}k_\perp^{\mu_1} k_\perp^{\mu_2} \dots k_\perp^{\mu_r}
	\exp \bigg(\frac{\kp^2+\pp^2}{|eB|}\bigg)   \exp\FB{\frac{2t}{1-t} \frac{\kp^2}{|eB|} + \frac{2u}{1-u}\frac{\pp^2}{|eB|}}  \Big|_{\pp=\qp+\kp}. \label{I3}
\end{eqnarray}
Let us now shift integration variable $\kp \to (\kp-\qtil)$ where
\begin{eqnarray}
	\qtil = \frac{(1-t)(1+u)}{2(1-tu)} \qp \label{qtil}
\end{eqnarray}
so that Eq.~\eqref{I3} simplifies to  
\begin{eqnarray}
\sum_{l,n=0}^{\infty} I_{abln}^{\mu_1\dots\mu_r}t^lu^n &&= 
	\frac{ \exp\TB{-\xi\frac{(1+t)(1+u)}{(1-tu)}}}{(1-t)^{a+1}(1-u)^{b+1}}
\int\!\!\dfrac{d^2k_\perp}{(2\pi)^2} \exp\TB{\frac{(1-tu)}{(1-t)(1-u)} \frac{2\kp^2}{|eB|}} 
 (k_\perp^{\mu_1}-\qtil^{\mu_1}) (k_\perp^{\mu_2}-\qtil^{\mu_2}) \dots (k_\perp^{\mu_r}-\qtil^{\mu_r}) \nn \\  
 && \hspace{-1cm}=  \frac{ \exp\TB{-\xi\frac{(1+t)(1+u)}{(1-tu)}}}{(1-t)^{a+1}(1-u)^{b+1}}
\int\!\!\dfrac{d^2k_\perp}{(2\pi)^2} \exp\TB{\frac{(1-tu)}{(1-t)(1-u)} \frac{2\kp^2}{|eB|}} 
(-1)^r \AB{\qtil^{\mu_1}\dots\qtil^{\mu_r} + \kp^{\mu_1}\kp^{\mu_2} \qtil^{\mu_3}\dots\qtil^{\mu_r} 
	+ \dots}
\label{I4}
\end{eqnarray}
where $\xi = \dfrac{\qpv^2}{2|eB|}\ge0$, and in the last step we have dropped the terms having odd powers in $\kp$ from the integrand. The angular bracket $\AB{A^{\mu_1\mu_2\dots\mu_r}}$ appearing in Eq.~\eqref{I4} denotes sum of all the \emph{distinct} terms generated from all possible permutations of the Lorentz indices. Eq.~\eqref{I4} can be rewritten in a compact form as follows 
\begin{eqnarray}
	\sum_{l,n=0}^{\infty} I_{abln}^{\mu_1\dots\mu_r}t^lu^n &=& 	
	\frac{ \exp\TB{-\xi\frac{(1+t)(1+u)}{(1-tu)}}}{(1-t)^{a+1}(1-u)^{b+1}}
	\int\!\!\dfrac{d^2k_\perp}{(2\pi)^2} \exp\TB{\frac{(1-tu)}{(1-t)(1-u)} \frac{2\kp^2}{|eB|}}  (-1)^r \sum_{s=0}^{\floor{r/2}} \AB{ \kp^{\mu_1}\dots\kp^{\mu_{2s}} \qtil^{\mu_{2s+1}}\dots\qtil^{\mu_r} } \label{I5}
\end{eqnarray}
where $\floor{x}$ is the floor function giving largest integer less than or equal to $x$. Using the symmetry arguments~\cite{Peskin:1995ev}, we can replace $\kp^{\mu_1}\kp^{\mu_2}\dots \kp^{\mu_{2s}} \to \dfrac{(\kp^2)^s}{(2s)!!}\AB{g_\perp^{\mu_1\mu_2} \dots g_\perp^{\mu_{2s-1}\mu_{2s} }}$ where $(2s)!!=2^ss!$, so that Eq.~\eqref{I5} becomes
\begin{eqnarray}
	\sum_{l,n=0}^{\infty}\!\! I_{abln}^{\mu_1\dots\mu_r}t^lu^n \!\!&=& 	\!\!(-1)^r \!\!\sum_{s=0}^{\floor{r/2}}
	\frac{ \exp\!\TB{-\xi\frac{(1+t)(1+u)}{(1-tu)}}}{(1-t)^{a+1}(1-u)^{b+1}}
	\!\int\!\!\dfrac{d^2k_\perp}{(2\pi)^2} \!\exp\!\TB{\frac{(1-tu)}{(1-t)(1-u)} \frac{2\kp^2}{|eB|}}  \! \frac{(\kp^2)^s}{2^ss!} \!
	\AB{ g_\perp^{\mu_1\mu_2}\!\!\!\dots g_\perp^{\mu_{2s-1}\mu_{2s}} \qtil^{\mu_{2s+1}}\!\!\!\dots\qtil^{\mu_r} \!}. \label{I6}
\end{eqnarray}
We now note that the integral on the rhs of Eq.~\eqref{I6} can trivially be performed, to obtain the following expression after substitution of $\qtil$ from Eq.~\eqref{qtil}
\begin{eqnarray}
	\sum_{l,n=0}^{\infty} I_{abln}^{\mu_1\dots\mu_r}t^lu^n \!&=& \!\!	 \frac{|eB|}{8\pi} \sum_{s=0}^{\floor{r/2}} \frac{(-1)^{r+s} }{2^{r} }  |eB|^s
	\AB{ g_\perp^{\mu_1\mu_2}\dots g_\perp^{\mu_{2s-1}\mu_{2s}} \qp^{\mu_{2s+1}}\dots\qp^{\mu_r} } 
	\exp\!\TB{-\xi\frac{(1+t)(1+u)}{(1-tu)}} 
		\frac{(1-tu)^{-1-r+s}(1+u)^{r-2s}}{(1-t)^{a-r+s}(1-u)^{b-s}} 
	\nn \\
		\!&=& \sum_{s=0}^{\floor{r/2}} T^{\mu_1\dots\mu_r}_s 
		\exp\TB{-\xi-\xi\frac{t(1+u)}{(1-tu)} -\xi\frac{u(1+t)}{(1-tu)}} 				\frac{(1-tu)^{-1-r+s}(1+u)^{r-2s}}{(1-t)^{a-r+s}(1-u)^{b-s}} 
	\label{I7} 
\end{eqnarray}
where 
\begin{eqnarray}
T^{\mu_1\dots\mu_r}_s = \dfrac{|eB|}{8\pi}\dfrac{(-1)^{r+s}}{2^r} |eB|^s \AB{ g_\perp^{\mu_1\mu_2}\dots g_\perp^{\mu_{2s-1}\mu_{2s}} \qp^{\mu_{2s+1}}\dots\qp^{\mu_r} }. \label{T}
\end{eqnarray}

Let us now use the following generating function of generalized Laguerre polynomials~\cite{GRADSHTEYN1980904} 
\begin{eqnarray}
	e^{-z\xi} = \frac{1}{(1+z)^\alpha} \sum_{l=0}^{\infty} L_l^{\alpha-l}(\xi)z^l ~~~\text{for}~ |z|<1	
\end{eqnarray} 
to replace the exponential factors in the rhs of Eq.~\eqref{I7} to obtain 
\begin{eqnarray}
	\sum_{l,n=0}^{\infty}\!  I_{abln}^{\mu_1\dots\mu_r}t^lu^n &=&\!  \sum_{s=0}^{\floor{r/2}}  T_s^{\mu_1\dots\mu_r}	\frac{(1-tu)^{-1-r+s}(1+u)^{r-2s}}{(1-t)^{a-r+s}(1-u)^{b-s}} 
e^{-\xi}  \FB{\frac{1-tu}{1+t}}^{\alpha} \FB{\frac{1-tu}{1+u}}^{\beta} \!\sum_{l,n=0}^{\infty}\! L_l^{\alpha-l}(\xi)\FB{\frac{t+tu}{1-tu}}^l  L_n^{\beta-n}(\xi)\FB{\frac{u+tu}{1-tu}}^n \nn \\ 
&=&\!  e^{-\xi} \sum_{s=0}^{\floor{r/2}}  T_s^{\mu_1\dots\mu_r} \!\sum_{l,n=0}^{\infty}\!  L_l^{\alpha-l}(\xi) L_n^{\beta-n}(\xi) t^lu^n 
\frac{(1-t)^{-a+r-s}(1-u)^{-b+s}(1+t)^{-\alpha+n}(1+u)^{-\beta+l+r-2s}}{(1-tu)^{-\alpha+l-\beta+n+1+r-s}}
\label{I8}
\end{eqnarray}
where the $\alpha$ and $\beta$ will be chosen later as per our convenience. Using Binomial series expansion $ (1+x)^a = \sum_{i=0}^{\infty} \binom{a}{i}x^i$ for $|x| < 1$, we can rewrite Eq.~\eqref{I8} as
\begin{eqnarray}
	\sum_{l,n=0}^{\infty}  I_{abln}^{\mu_1\dots\mu_r}t^lu^n &=&  e^{-\xi} \sum_{s=0}^{\floor{r/2}}  T_s^{\mu_1\dots\mu_r} \sum_{l,n=0}^{\infty} ~~\sum_{i,j,i',j',m=0}^{\infty} L_l^{\alpha-l}(\xi) L_n^{\beta-n}(\xi) (-1)^{i+j+m} t^{l+i+i'+m} u^{n+j+j'+m} 
	\nn \\ && \times \binom{-a+r-s}{i} \binom{-b+s}{j} \binom{-\alpha+n}{i'} \binom{-\beta+l+r-2s}{j'} \binom{\alpha-l+\beta-n-1-r+s}{m}.  \label{I9}
\end{eqnarray}
From Eq.~\eqref{I9}, we can now extract $ I_{abln}^{\mu_1\dots\mu_r}$ as 
\begin{eqnarray}
	 I_{abln}^{\mu_1\dots\mu_r} &=& \frac{1}{l!n!}  \FB{\frac{\del^l}{\del t^l}} \FB{\frac{\del^n}{\del u^n}} \sum_{l',n'=0}^{\infty}  I_{abl'n'}^{\mu_1\dots\mu_r}t^{l'}u^{n'}\bigg|_{t=u=0} \nn \\
	 &=& e^{-\xi} \sum_{s=0}^{\floor{r/2}}  T_s^{\mu_1\dots\mu_r} \sum_{l',n'=0}^{\infty} ~~\sum_{i,j,i',j',m=0}^{\infty} L_{l'}^{\alpha-l'}(\xi) L_{n'}^{\beta-n'}(\xi) (-1)^{i+j+m} \delta_l^{l'+i+i'+m} \delta_n^{n'+j+j'+m} 
	 \nn \\ && \times \binom{-a+r-s}{i} \binom{-b+s}{j} \binom{-\alpha+n'}{i'} \binom{-\beta+l'+r-2s}{j'} \binom{\alpha-l'+\beta-n'-1-r+s}{m} \label{I10}
\end{eqnarray}
where, $\delta_i^j$ is the Kronecker delta function. Using these $\delta$-functions, we evaluate the sum over indices $l'$ and $n'$ on the rhs of Eq.~\eqref{I10} to obtain
\begin{eqnarray}
	I_{abln}^{\mu_1\dots\mu_r} \!&=& e^{-\xi} \sum_{s=0}^{\floor{r/2}}  T_s^{\mu_1\dots\mu_r}  \sum_{i,j,i',j',m=0}^{\infty} L_{l-i-i'-m}^{\alpha-l+i+i'+m}(\xi) L_{n-j-j'-m}^{\beta-n+j+j'+m}(\xi) (-1)^{i+j+m}
	 \binom{-a+r-s}{i} \binom{-b+s}{j} \binom{-\alpha+n-j-j'-m}{i'} 
	 \nn \\ && \hspace{-1cm} \times
	 \binom{-\beta+l-i-i'-m+r-2s}{j'} \binom{\alpha-l+i+i'+m+\beta-n+j+j'+m-1-r+s}{m}  \theta(l-i-i'-m) \theta(n-j-j'-m)\label{I11}
\end{eqnarray}
where $\theta(x)$ is the unit step function. We now choose $\beta = (l-i-i'-m+r-2s)$ so that $\binom{-\beta+l-i-i'-m+r-2s}{j'} = \binom{0}{j'} = \delta_{j'}^0$, which in turn enable us to evaluate the sum over index $j'$. Hence, Eq.~\eqref{I11} simplifies to the following
\begin{eqnarray}
	I_{abln}^{\mu_1\dots\mu_r} \!&=& e^{-\xi} \sum_{s=0}^{\floor{r/2}}  T_s^{\mu_1\dots\mu_r}  \sum_{i,j,i',m=0}^{\infty} L_{l-i-i'-m}^{\alpha-l+i+i'+m}(\xi) L_{n-j-m}^{l-i-i'+r-2s-n+j}(\xi) (-1)^{i+j+m}
\binom{-a+r-s}{i} \binom{-b+s}{j} \binom{-\alpha+n-j-m}{i'} 
\nn \\ &&  \times
 \binom{\alpha-n+j-s+m-1}{m}  \theta(l-i-i'-m) \theta(n-j-m) \label{I12}
\end{eqnarray}
Let us now choose $\alpha=(n-j+s)$ so that $\binom{\alpha-n+j-s+m-1}{m} = \binom{m-1}{m} = \delta_{m}^0$, which further enable us to evaluate the sum over index $m$ and we obtain from Eq.~\eqref{I12} the following
\begin{eqnarray}
	I_{abln}^{\mu_1\dots\mu_r} &=& e^{-\xi} \sum_{s=0}^{\floor{r/2}}  T_s^{\mu_1\dots\mu_r}  \!\!\!\sum_{i,j,i'=0}^{\infty}\!\!\! L_{l-i-i'}^{n-j+s-l+i+i'}(\xi) L_{n-j}^{l-i-i'+r-2s-n+j}(\xi) (-1)^{i+j}
	\binom{-a+r-s}{i} \binom{-b+s}{j} \binom{-s}{i'} 
	  \theta(l-i-i') \theta(n-j) \nn \\
	  &=& e^{-\xi} \sum_{s=0}^{\floor{r/2}}  T_s^{\mu_1\dots\mu_r}  \sum_{i=0}^{l} \sum_{j=0}^{n} \sum_{i'=0}^{l-i}
	  L_{l-i-i'}^{n-j+s-l+i+i'}(\xi) L_{n-j}^{l-i-i'+r-2s-n+j}(\xi) (-1)^{i+j}
	  \binom{-a+r-s}{i} \binom{-b+s}{j} \binom{-s}{i'} \label{I13}
\end{eqnarray}
where, in the last step, the unit step functions are used to restrict the upper limit of the summation over indices $i,j,i'$. Now replacing $i' \to (l-i-i')$ in the summand of Eq.~\eqref{I13} yields
\begin{eqnarray}
	I_{abln}^{\mu_1\dots\mu_r} &=& e^{-\xi} \sum_{s=0}^{\floor{r/2}}  T_s^{\mu_1\dots\mu_r}  \sum_{i=0}^{l} \sum_{j=0}^{n} \sum_{i'=0}^{l-i}
	L_{i'}^{n-j+s-i'}(\xi) L_{n-j}^{i'+r-2s-n+j}(\xi) (-1)^{i+j}
	\binom{-a+r-s}{i} \binom{-b+s}{j} \binom{-s}{l-i-i'} \label{I14}
\end{eqnarray}
which further simplifies upon replacing $i\to(l-i)$ and $j\to(n-j)$ into
\begin{eqnarray}
	I_{abln}^{\mu_1\dots\mu_r} &=& (-1)^{l+n} e^{-\xi} \sum_{s=0}^{\floor{r/2}}  T_s^{\mu_1\dots\mu_r}  \sum_{i=0}^{l} \sum_{j=0}^{n} \sum_{i'=0}^{i}
	L_{i'}^{j+s-i'}(\xi) L_{j}^{i'+r-2s-j}(\xi) (-1)^{i+j}
	\binom{-a+r-s}{l-i} \binom{-b+s}{n-j} \binom{-s}{i-i'}. \label{I15}
\end{eqnarray}
 We now substitute $T_s^{\mu_1\dots\mu_r}$ from Eq.~\eqref{T} into Eq.~\eqref{I15} to obtain the final expression 
\begin{eqnarray}
	%
I_{abln}^{\mu_1\dots\mu_r}(\qp,|eB|)	&=& (-1)^{l+n} \frac{|eB|}{8\pi} \sum_{s=0}^{\floor{r/2}} |eB|^s \AB{ g_\perp^{\mu_1\mu_2}\dots g_\perp^{\mu_{2s-1}\mu_{2s}} \qp^{\mu_{2s+1}}\dots\qp^{\mu_r} } \mathcal{A}_{abln}^{rs}(\xi) \label{I16}
\end{eqnarray}
where we recall $\xi = \frac{\qpv^2}{2|eB|}\ge0$ and the dimensionless quantity $\mathcal{A}_{abln}^{rs}(\xi)$ is given by
\begin{eqnarray}
	\mathcal{A}_{abln}^{rs}(\xi) =  e^{-\xi}   \dfrac{(-1)^{r+s}}{2^r} 
 \sum_{i=0}^{l} \sum_{j=0}^{n} \sum_{i'=0}^{i}
	L_{i'}^{j+s-i'}(\xi) L_{j}^{i'+r-2s-j}(\xi) (-1)^{i+j}
	\binom{-a+r-s}{l-i} \binom{-b+s}{n-j} \binom{-s}{i-i'}.	 \label{A1}
\end{eqnarray}
Eq.~\eqref{I16} is the final analytic expression for the integral in Eq.~\eqref{I1} for an arbitrary $r$ in terms of finite sums. In the next section, we will provide explicit expressions for $r=0,1,2,3$ and 4 which appear in most of the quantum field theoretical calculations. 

\section{Explicit Expressions of the integral up to Fourth Moment} \label{sec.moments}
Putting $r=0,1,2,3$ and 4 successively in Eq.~\eqref{I16} yields the following expressions 
\begin{eqnarray}
	I_{abln} &=& (-1)^{l+n}\frac{|eB|}{8\pi} \mathcal{A}_{abln}^{00}(\xi), \\ 
	I_{abln}^{\mu_1} &=& (-1)^{l+n}\frac{|eB|}{8\pi} \qp^{\mu_1} \mathcal{A}_{abln}^{10}(\xi), \\ 
	I_{abln}^{\mu_1\mu_2} &=& (-1)^{l+n}\frac{|eB|}{8\pi} \Big[  \qp^{\mu_1} \qp^{\mu_2} \mathcal{A}_{abln}^{20}(\xi) + |eB|g_\perp^{\mu_1\mu_2} \mathcal{A}_{abln}^{21}(\xi)  \Big], \\ 
	I_{abln}^{\mu_1\mu_2\mu_3} &=& (-1)^{l+n}\frac{|eB|}{8\pi} \Big[  \qp^{\mu_1} \qp^{\mu_2} \qp^{\mu_3} \mathcal{A}_{abln}^{30}(\xi) + |eB| \AB{g_\perp^{\mu_1\mu_2}\qp^{\mu_3}} \mathcal{A}_{abln}^{31}(\xi)  \Big] \nn \\
	&=& (-1)^{l+n}\frac{|eB|}{8\pi} \Big[  \qp^{\mu_1} \qp^{\mu_2} \qp^{\mu_3} \mathcal{A}_{abln}^{30}(\xi) + |eB| \Big(g_\perp^{\mu_1\mu_2}\qp^{\mu_3} + g_\perp^{\mu_1\mu_3}\qp^{\mu_2} + g_\perp^{\mu_2\mu_3}\qp^{\mu_1} \Big) \mathcal{A}_{abln}^{31}(\xi)  \Big], 	\\
	I_{abln}^{\mu_1\mu_2\mu_3\mu_4} &=& (-1)^{l+n}\frac{|eB|}{8\pi} \Big[  \qp^{\mu_1} \qp^{\mu_2} \qp^{\mu_3}\qp^{\mu_4} \mathcal{A}_{abln}^{40}(\xi) + |eB| \AB{g_\perp^{\mu_1\mu_2}\qp^{\mu_3}\qp^{\mu_4}} \mathcal{A}_{abln}^{41}(\xi) + |eB|^2 \AB{g_\perp^{\mu_1\mu_2} g_\perp^{\mu_3\mu_4} } \mathcal{A}_{abln}^{42}(\xi) \Big] \nn \\	
	&=& (-1)^{l+n}\frac{|eB|}{8\pi} \Big[  \qp^{\mu_1} \qp^{\mu_2} \qp^{\mu_3}\qp^{\mu_4} \mathcal{A}_{abln}^{40}(\xi) + |eB| \Big(g_\perp^{\mu_1\mu_2}\qp^{\mu_3}\qp^{\mu_4} + g_\perp^{\mu_1\mu_3}\qp^{\mu_2}\qp^{\mu_4} + g_\perp^{\mu_1\mu_4}\qp^{\mu_2}\qp^{\mu_3} + g_\perp^{\mu_2\mu_3}\qp^{\mu_1}\qp^{\mu_4} \nn \\
	&& \hspace{.6cm} +~g_\perp^{\mu_2\mu_4}\qp^{\mu_1}\qp^{\mu_3} + g_\perp^{\mu_3\mu_4}\qp^{\mu_1}\qp^{\mu_2}\Big) \mathcal{A}_{abln}^{41}(\xi) 
	+ |eB|^2 \Big(g_\perp^{\mu_1\mu_2} g_\perp^{\mu_3\mu_4} + g_\perp^{\mu_1\mu_3} g_\perp^{\mu_2\mu_4} + g_\perp^{\mu_1\mu_4} g_\perp^{\mu_2\mu_3}\Big) \mathcal{A}_{abln}^{42}(\xi) \Big].
\end{eqnarray}

We end this section by mentioning two useful contractions to the third and fourth moments which may be required in field theoretical calculations:
\begin{eqnarray}
	g_{\perp \mu_1\mu_2}q_{\perp\mu_3} 	I_{abln}^{\mu_1\mu_2\mu_3} &=& (-1)^{l+n}\frac{|eB|}{8\pi} \qp^2 \Big[ \qp^2 \mathcal{A}_{abln}^{30}(\xi) + 4 |eB| \mathcal{A}_{abln}^{31}(\xi)  \Big] , \\
	g_{\perp \mu_1\mu_2}	g_{\perp \mu_3\mu_4} 	I_{abln}^{\mu_1\mu_2\mu_3\mu_4} &=& (-1)^{l+n}\frac{|eB|}{8\pi} \Big[  \qp^4 \mathcal{A}_{abln}^{40}(\xi) + 8 |eB| \qp^2   \mathcal{A}_{abln}^{41}(\xi) + 8 |eB|^2  \mathcal{A}_{abln}^{42}(\xi) \Big].
\end{eqnarray}
%
%

\section*{Acknowledgments}
The author thanks Dr. Arunava Maity and Dr. Sonia Mandal for useful discussions. The author   acknowledges Prof. Sourav Sarkar for immense encouragement and support in all respect. S.G. is funded by the Department of Higher Education, Government of West Bengal, India.


\bibliography{z-Ref}

\end{document}